\documentclass[showpacs,twocolumn,aps,pre,floatfix,noshowpacs]{revtex4-1}
\usepackage{amsthm}
\usepackage[mathlines]{lineno}
\usepackage[normalem]{ulem}
\usepackage{etoolbox}
\usepackage{amsmath,amssymb,graphicx,bm,color,mathrsfs,verbatim,epstopdf,dcolumn,hyperref,tikz}

\begin{document}

\title{Dynamic scaling of the restoration of rotational symmetry \\ in Heisenberg quantum antiferromagnets}

\author{Phillip Weinberg and Anders W. Sandvik}
\affiliation{Department of Physics, Boston University, 590 Commonwealth Avenue, Boston, Massachusetts 02215, USA}

\begin{abstract}
We apply imaginary-time evolution with the operator ${\rm e}^{-\tau H}$ to study relaxation dynamics of gapless quantum antiferromagnets described by the spin-rotation invariant Heisenberg Hamiltonian ($H$). Using quantum Monte Carlo simulations to obtain unbiased results, we propagate an initial state with maximal order parameter $m^z_s$ (the staggered magnetization) in the $z$ spin direction and monitor the expectation value $\langle m_s\rangle $ as a function of imaginary-time $\tau$. Results for different system sizes (lengths) $L$ exhibit an initial essentially size-independent relaxation of $\langle m_s\rangle$ toward its value in the infinite-size spontaneously symmetry-broken state, followed by a strongly size-dependent final decay to zero when the O$(3)$ rotational symmetry of the order paraneter is restored. We develop a generic finite-size scaling theory that shows the relaxation time diverges asymptotically as $L^z$ where $z$ is the dynamic exponent of the low energy excitations. We use the scaling theory to develop a practical way of extracting the dynamic exponent from the numerical finite-size data, systematcally eliminating scaling corrections. We apply the method to spin-$1/2$ Heisenberg antiferromagnets on two different lattice geometries: the standard two-dimensional (2D) square lattice as well as a site-diluted 2D square lattice at the percolation threshold. In the 2D case we obtain $z=2.001(5)$, which is consistent with the known value $z=2$, while for the site-dilutes lattice we find $z=3.90(1)$ or $z=2.056(8)D_f$, where $D_f=91/48$ is the fractal dimensionality of the percolating system. This is an improvement on previous estimates of $z\approx 3.7$. The scaling results also show a fundamental difference between the two cases; for the 2D square lattice, the data can be collapsed onto a common scaling function even when $\langle m_s\rangle$ is relatively large, reflecting the Anderson tower of quantum rotor states with a common dynamic exponent $z=2$. For the diluted 2D square lattice, the scaling works well only for small $\langle m_s\rangle$, indicating a mixture of different relaxation time scaling between the low energy states. Nevertheless, the low-energy dynamic here also corresponds to a tower of excitations.
\end{abstract}

\maketitle
\section{\label{sec:intro} Introduction}

Experimental studies of interacting quantum systems are increasingly focusing on non-equilibrium setups, e.g., driving cold atom systems or electronic materials dynamically through various finite temperature and quantum phase transitions \cite{paillard01,kaiser14,loth10,bloch05,sadler06,ritter07,trotzky11,schreiber15}. Theoretical modeling of systems under these conditions is even more challenging than the already difficult problem of computing equilibrium properties of quantum systems away from perturbative regimes. Exact numerical calculations are possible for small systems, and there has been some success in studying issues such as thermalization and many-body localization~\cite{nandkishore15,vosk13,pal10,oganesyan07,imbrie16,gornyi05,baskoa06}. Reaching system sizes sufficiently large enough for modeling experiments is still difficult in most cases, with the exception of some one-dimensional (1D) systems where the density-matrix renormalization group (DMRG) method (or the closely related matrix-product states) now allows for time-evolution studies on relatively large system sizes and long times \cite{daley04,schollwock05,schollwock11}. For higher-dimensional systems the challenges in real-time calculations are formidable.

Given the difficulties with real-time evolution, alternative ways to extract non-equilibrium dynamical properties of quantum systems have been explored in the imaginary-time domain, where quantum Monte Carlo (QMC) methods can be applied. In equilibrium, there are solid relationships between real- and imaginary-time correlation functions which can be exploited in numerical analytic continuation of QMC data~\cite{jarrell96}. Much less is known about practical ways to infer real-time properties from imaginary-time calculations out of equilibrium, though some progress has been made on this front recently. Examples include studies of systems driven through a quantum critical point \cite{degrandi11,degrandi13,liu13} at different velocities or according to non-linear protocols, where Kibble-Zurek scaling \cite{kibble76,zurek85,polkovnikov05,zurek05,dziarmaga05,zhong05,chandran12,liu14} can be used to extract the dynamic exponent and other important quantities such as the quantum geometric tensor \cite{micheal14,degrandi11,degrandi13}. An important observation here is that real- and imaginary-time evolution are identical not only in the adiabatic limit, where both dynamics keep systems in their instantaneous ground states, but also including the leading non-adiabatic effects. Another example is the phenomenon of ``initial slip'' \cite{yin14,zhang14,shu17,chiocchetta16,liu16}, where a random product state is evolved by a Hamiltonian tuned to a quantum-critical point and the state initially becomes increasingly ordered, before developing critical fluctuations and vanishing long-range order. The transient states produced before one reaches the critical equilibrium state have interesting properties that can be probed in imaginary-time \cite{yin14,zhang14,shu17}. Imaginary-time evolution has also been used to investigate the emergence of topological conservation laws \cite{shao15}.

In this paper we use the imaginary-time approach to study the relaxation mechanism of the order parameter in quantum antiferromagnets with O($3$) rotationally-invariant order parameters. The system, described by a Hamiltonian $H$, is initially prepared in a fully saturated antiferromagnetic state with the order parameter along the $z$ spin axis. Evolving this state in imaginary-time $\tau$ with the operator ${\rm e}^{-\tau H}$, the rotational symmetry will eventually be restored marked by the expectation value of the $z$ component of the order parameter decaying to zero. We identify short- and long-time behavior of the dynamics and develop theoretical and practical tools for analyzing emergent scaling behaviors by defining an effective dynamic exponent for a given threshold value of the $z$ component of the order parameter. As we shall show in this paper, this effective dynamic exponent converges to the true dynamic exponent as the threshold of the order parameter is lowered to $0$. In addition to delivering the dynamic exponent, which is not very surprising (though useful in its utility as a tool to extract the exponent), we show that the asymptotic long-time, large-system scaling behavior contains valuable information on the nature of the high-energy states. In a clean 2D Heisenberg antiferromagnet we observe fast convergence of the effective dynamic as the threshold order parameter is decreased, which we argue is indicative of the expected ``Anderson tower'' of quantum rotor states \cite{anderson52}. In other words, the scaling is characterized by an constant effective dynamic exponent $z=D$ ($D$ being the dimensionality of the system) for a large range of values of the order parameter. In contrast, in a 2D system randomly diluted at its percolation point, we find that the effective dynamic exponent increases as the threshold order parameter goes to 0, converging toward a fixed value only when the order parameter is small. This demonstrates a hierarchy of excitations which forms tower governed by a common dynamic (size-scaling) exponent only at low energies. The ultimate low-energy value of the dynamic exponent is $z = 3.90(1)$, which improves in previous estimates $z \approx 3.7$ obtained using different methods \cite{wang06,wang10,ghosh15}. These results reinforce the notion that lowest excitations of the system at the percolation point are not conventional Anderson quantum rotor states (Goldstone modes \cite{anderson52,goldstone62}) although the system breaks the $O(3)$ spin symmetry spontaneously in the thermodynamic limit \cite{sandvik02,vojta95}.

The outline of the rest of the paper is as follows:
In Sec.~\ref{sec:theory} we will describe the theoretical underpinnings of our approach. In Secs.~\ref{sec:cleanheisenberg} and \ref{sec:percolating} we
discuss results for the pure 2D Heisenberg model and the diluted system, respectively. We summarize our study and provide some further remarks
in Sec.~\ref{summary}.

\section{\label{sec:theory}Relaxation and Finite Size Scaling}

As mentioned in the introduction our setup will be the following: We prepare our system initially in a fully saturated antiferromagnetic state denoted by $|\psi_0\rangle$, and evolve in imaginary time with $\mathrm{e}^{-\tau H}$ where $H$ is a Heisenberg Hamiltonian describing a gapless quantum antiferromagnet.  Because the ground state obeys the symmetry of $H$ we know that this initial state will not be an eigenstate of $H$ but it will have some overlap with the ground state, as they share the same ordering. As the state evolves in imaginary-time it will eventually decay to the ground state, restoring the rotational symmetry of our system. From the theory of spontaneous symmetry breaking we know that in the limit of system size tending to infinity, a set of excited states just above the rotationally symmetric ground state become degenerate allowing the system to spontaneously align along a particular axis when subjected to an infinitesimal perturbation \cite{d_fisher89,goldstone62}. In imaginary-time this phenomena will manifest itself as a divergence in the relaxation time to reach the ground state as the system size is increased. We will use the staggered magnetization $m_s^z$ along the $z$ spin axis
\begin{equation}
m_s^z = \sum_{i_x,i_y} (-1)^{i_x+i_y} S^z_{i_x,i_y},
\end{equation}
as a measure of the restoration of rotational symmetry as our state is evolving in imaginary-time. The expectation value of this operator as a function of imaginary-time is given by:
\begin{equation}
\langle m_s^z(\tau) \rangle=\frac{\langle\psi(\tau)|m_s^z|\psi(\tau)\rangle}{\langle\psi(\tau)|\psi(\tau)\rangle},\label{eq:exp}
\end{equation}
where $|\psi(\tau)\rangle$ is the imaginary-time evolved state:
\begin{equation}
|\psi(\tau)\rangle=\exp\left(-\tau H\right)|\psi(0)\rangle.
\end{equation}
Expanding in eigenstates of $H$, denoting the eignestates and eigenenergies by $|n\rangle$ and $\epsilon_n$, respectively, and defining the gap, 
$\Delta=\epsilon_1-\epsilon_0$; in the limit $\tau\rightarrow\infty$, the expectation value in Eq.~(\ref{eq:exp}) will vanish as
\begin{equation}
\langle m_s^z(\tau)\rangle \approx\left(\langle 0|m_s^z|1\rangle \frac{c_{1}}{c_{0}} +~\mathrm{c.}~\mathrm{c.}\right)e^{-\Delta\tau} + \cdots,\label{eq:asm_imag_time}
\end{equation}
where $c_n=\langle n|\psi_0\rangle$. Since the ground state is symmetric under rotations we have that $\langle 0|m_s^z|0\rangle=0$. let us define the relaxation time $\tau_r$ as the time where $m_s^z(\tau)$ drops below some threshold $m_\mathrm{threshold}$. Equation~\eqref{eq:asm_imag_time} suggests that as this threshold goes to 0, $\tau_r \sim 1/\Delta$. Therefore in this limit, by calculating the scaling $\tau_r$ with system size, we can infer the scaling of the low energy gap of $H$. 

One can characterize this scaling of the low energy gap by the dynamic exponent $\Delta\sim L^{-z}$ which has different interpretations depending on its value. $z=0$ implies that the system has a finite gap in the thermodynamic limit, while finite $z$ means the system has gapless excitations, and finally $z=\infty$ denotes exponential scaling of the gap. If we consider systems which have gapless excitations, a finite dynamic exponent implies the relaxation time should scale as a power law: $\tau_r\sim L^z$. In other words, if one rescales the time axis by $L^z$, $\langle m_s^z(\tau)\rangle$ should show scaling collapse at small $m_\mathrm{threshold}$. For some Hamiltonians there exists a large, but sub-extensive, number of low energy states which have energies (relative to the ground state) that have the same scaling as the low energy gap~\cite{anderson52,goldstone62}. The existence of these states will have the effect that there will be a larger window of $m_\mathrm{threshold}$ for which this scaling collapse holds. This is because higher order terms in Eq.~(\eqref{eq:asm_imag_time}) will have energy exponentials which all scale in a similar manner with system size. This argument will be important later when we discuss the differences between the clean and diluted 2D Heisenberg models in later sections.

The preceding arguments, however, become valid asymptotically in the limit $L\rightarrow\infty$ and so it is necessary to take into account finite size corrections if one would like quantitative estimates for the dynamic exponent. As we will explain in the rest of this section, it is possible to control for the effects of finite size deviations by calculating the "flow" of the dynamic exponent from finite-size systems in a similar manner as the techniques used in finite-size scaling near critical points in equilibrium~\cite{nightingale82,brezin82,derrida83,luck85}. Using these methods one can extrapolating the finite-size results to the thermodynamic limit. 

To estimate the dynamic exponent from finite size systems we start by calculating the relaxation times $\tau_r$ and $\tau_r'$ (at some finite value of $m_\mathrm{threshold}$) for two different system sizes $L$ and $L'$ respectively. From this, the dynamic exponent can be estimated by rescaling the two times by their respective system sizes such that the rescaled results are equal:
\begin{equation}
\tau_r L^{z(L,L')}=\tau_r'L'^{z(L,L')}.\label{eq:dyn_z}
\end{equation}
From this expression we can define a finite size exponent for the pair of system sizes $z(L,L')$:
\begin{equation}
z(L,L')=\frac{\log(\tau_r/\tau_r')}{\log(L/L')}.\label{eq:z_L}
\end{equation}
As $L,L'\rightarrow\infty$, this finite size exponent will converge to what we shall call the effective dynamic exponent (recall that $m_\mathrm{threshold}$ is finite) which we will denote as $z_\infty$. The manner in which $z(L,L')$ converges to infinite size is determined by corrections to the $L^{-z_\infty}$ scaling of $\Delta(L)$. To see this, one can parametrize the corrections to the low energy gap with correction exponents $\omega_i$:
\begin{equation}
\Delta(L)=L^{-z_\infty}(1+c_1L^{-\omega_1}+\cdots).
\end{equation}
Now modifying Eq.~(\ref{eq:dyn_z}) to include corrections we get:
\begin{equation}
\tau_r \Delta(L)=\tau_r'\Delta(L').
\end{equation}
Using this equation and Eq.~(\eqref{eq:z_L}) we find that $z(L,L')$ converges to $z_\infty$ as:
\begin{equation}
z(L,L')=z_\infty+c_1\frac{L^{-\omega_1}-L'^{-\omega_1}}{\log(L'/L)}+\cdots.\label{eq:z_fit}
\end{equation}
The equation above allows one to extract both $\omega_i$'s and $c_i$'s needed to obtain the scaling corrections in $\Delta(L)$. Note that we have explicitly suppressed the fact that the effective dynamic exponent as well as the finite size corrections are functions of $m_\mathrm{threshold}$. We will revisit this functional dependence in later sections, but it is important to recall our previous argument that in the limit $m_\mathrm{threshold}\rightarrow 0$, $z_\infty$ converges to the true dynamic exponent of the model. 

In the following we will test this scaling hypotheses on two examples. First we study the Heisenberg antiferromagnetic on the simple square lattice. The low-energy physics of this model is very well understood and provides a good benchmark for our scaling approach~\cite{anderson52,d_fisher89,neuberger89,hasenfratz93}. We then go on to apply this method on the site-diluted Heisenberg antiferromagnet on a square lattice at the percolation point, where there have been previous studies but not as clear of a consensus as to the low energy physics of the model \cite{wang06,wang10,ghosh15}. To compute the imaginary-time evolution, we use a projector QMC method that in practice shares many similarities with the common Stochastic Series Expansion (SSE) method. We sample contributions to $\langle\psi_0|H^{2m}|\psi_0\rangle$ and define $\tau=m/N$ where $N$ is the total number of spins on the lattice. This time definition is equivalent, up to a factor, to the conventional imaginary-time appearing in the Schr\"odinger evolution operator~\cite{sandvik10,sandvik10_2}. 

\begin{figure}[h!]
	\includegraphics[width=7.5cm]{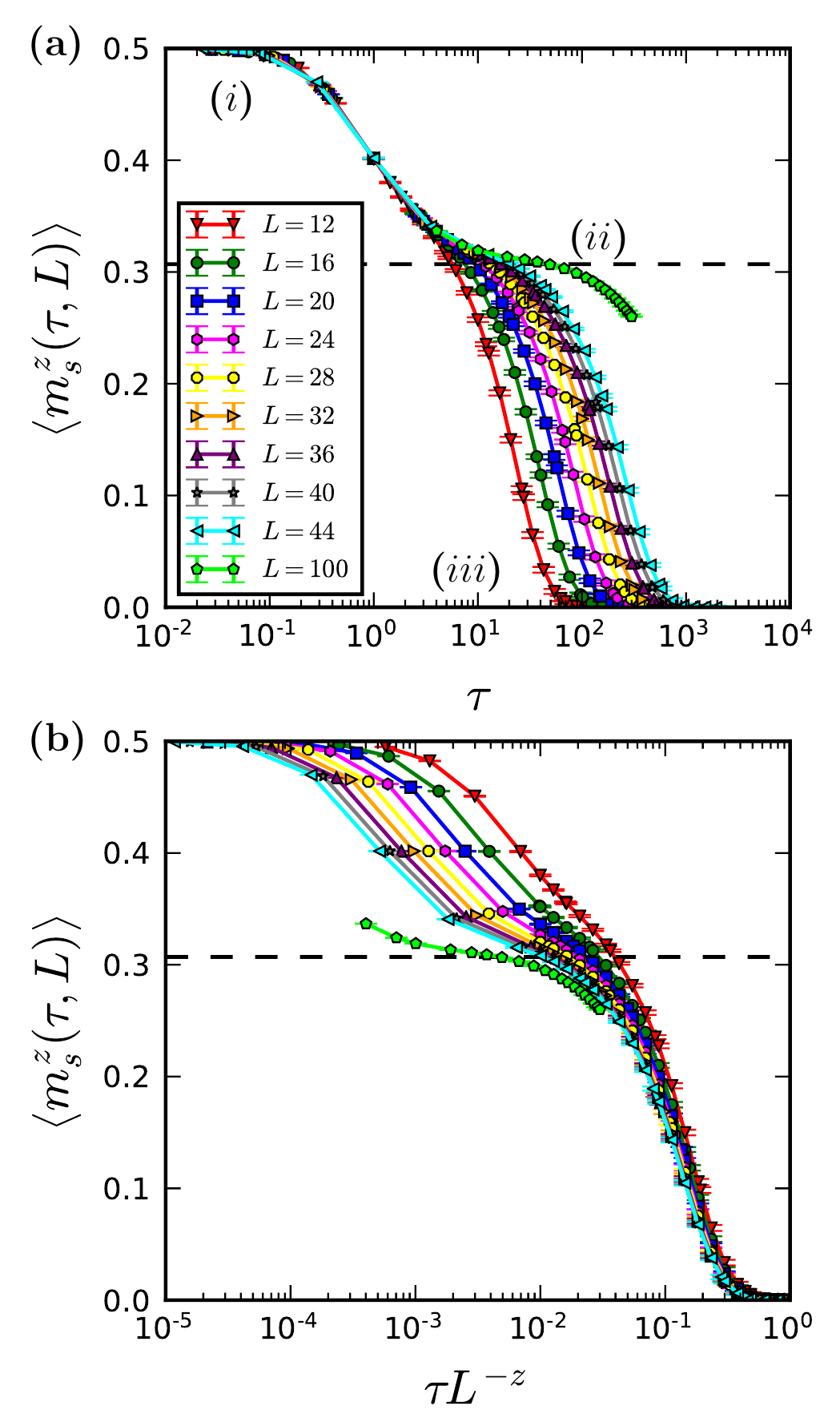}
	\caption{\label{fig:rotor}(a) Evolution of $\langle m_s^z\rangle $ as a function of imaginary-time in the clean 2D Heisenberg model for different system sizes. (b) The same data as in (a) but plotted vs $\tau L^{-z}$ ($z=D=2$), showing how the asymptotic relaxation is governed by the Anderson tower of rotor states. In both (a) and (b) the dotted line is the magnitude of the order parameter of the Heisenberg model in the thermodynamic limit at $T=0$. This is the value the order parameter would relax to if the system size was infinite and would remained in the symmetry broken ground state.}
\end{figure}

\begin{figure}[t]
	\includegraphics[width=0.48\textwidth]{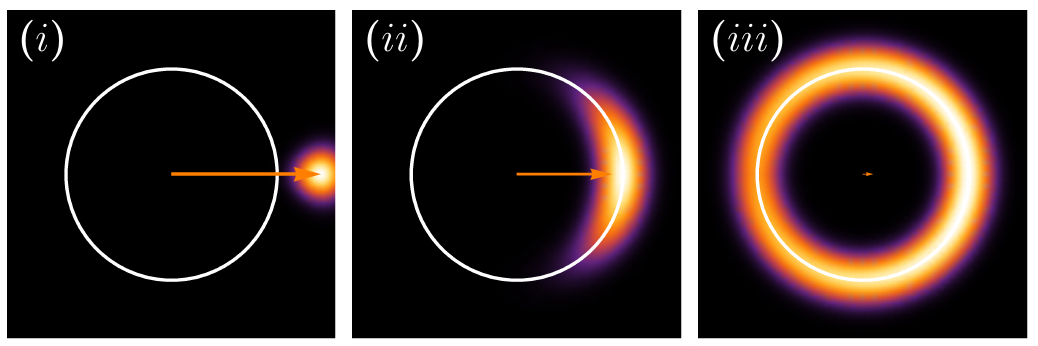}
	\caption{Pictural representations (not based on actual data) of the probability density of the order parameter in a 
	plane going through the origin of the O($3$) space. The panels (i)-(iii) correspond to the similarly marked scaling regimes in Fig.~\ref{fig:rotor}(a). The radius of the white circle corresponds to the magnitude of the order parameter in the ground state, and the arrow marks the average order parameter $\langle m_s^z\rangle$. In (i) the system is close to the initial state, where the magnitude of the order parameter is larger than in the ground state. In (ii) the magnitude has decayed to its asymptotic value but the direction of the order parameter is still close to the initial one. In (iii) the system is close to the asymptotic finite-size state with uniformly fluctuating (not symmetry-broken) order parameter.}
\label{fig:density}
\end{figure}

\section{Clean Heisenberg Antiferromagnet \label{sec:cleanheisenberg}}

For the first example we consider the standard two-dimensional (2D) antiferromagnetic Heisenberg model, defined by the Hamiltonian
\begin{equation}
H = J\sum_{\langle i,j\rangle} {\bf S}_i\cdot {\bf S}_j,
\end{equation}
where ${\bf S}_i$ is a spin-$1/2$ operator at lattice site $i$ and the sum is over nearest neighbors on an $L$ by $L$ square lattice in $D=2$ with periodic boundary conditions. We set $J=1$ for the rest of the paper. This model is a part of a broad class of antiferromagnet models on translationally-invariant bipartite lattices which are known to have quantum-rotor like low-energy excitations~\cite{anderson52}. These excited states have energy levels which become degenerate with the ground state in the limit $L\rightarrow\infty$ as a power law: $E_\mathrm{rotor}-E_\mathrm{GS}\sim L^{-z}$. Here $z=D$ is the dimension of the lattice and $L$ is the linear dimension of the system. One can think of these quantum rotor states as the quantization of the global angular fluctuations of the order parameter which in the thermodynamic limit form a basis in the ground state manifold used to create the symmetry broken ground state which is the vacuum of the gapless spinwave excitations \cite{d_fisher89,neuberger89,lavalle98,hasenfratz93}. 
Fig.~\ref{fig:rotor}(a) shows $\langle m^z_s(\tau)\rangle$ versus $\tau$ for various system sizes. In this figure one clearly sees that initially all the system sizes relax at the same rate but then eventually break off from one another when the sublattice magnetization reaches $\langle m^z_s(\tau)\rangle\approx 0.3$, with the larger system sizes taking longer to relax to $\langle m^z_s(\tau)\rangle\rightarrow 0$. One may recall that in the ground state of this model the thermodynamic value of $\langle m^z_s(T=0)\rangle\approx 0.307$ \cite{sandvik97} which roughly corresponds to the value where the different system sizes begin to relax at different rates (see the dotted line in Fig~\ref{fig:rotor}).

One can make use of the continuum field theory description of the long wavelength behavior of this model to understand the evolution of the order parameter in imaginary-time. The lowest energy states of the continuum field theory are uniform in space. Since the initial state is also uniform, and the system is translationally invariant, the evolution will occur within the subspace of uniform configurations. The dynamics in this subspace simplify to that of a quantum particle relaxing in the ``Mexican hat'' potential. Here the probability distribution of the particle in space represents the probability distribution of order parameter of the system. Since the initial order parameter value is $0.5$, the initial probability distribution of the order parameter is localized away from the minimum of the potential which for the Heisenberg model on a square lattice is close to $0.3$, as illustrated schematically in Fig.~\ref{fig:density}($i$). As the system evolves in imaginary-time the energy of the system decreases and order parameter decays until it reaches the bottom of the potential which in the thermodynamic limit corresponds to the the symmetry broken "ground state", as in Fig.~\ref{fig:density}($ii$). However, because the system is finite, this is not a true ground state so the energy of the system continues to decay and the mean magnetization along the z-axis relaxes to $0$ as the probability distribution spreads out over all possible solid angles; Fig.~\ref{fig:density}($iii$). This second part of the relaxation is governed by the rotor state as they are the states which make up the quantization of the angular part of the order parameter (and sets the effective moment of inertia which scales as $L^D$,~\cite{anderson52,sandvik10}). We can confirm this intuition for the lattice model by observing that in Fig.~\ref{fig:rotor}(b), by rescaling the $\tau$ axis with $L^2$ one finds that the second section of the relaxation shows scaling collapse. A consequence of this is that in the limit $L\rightarrow\infty$ the order parameter never relaxes to $0$ implying that the system remains in the symmetry broken ground state. 

Next let us discuss how to numerically extract the dynamic exponent from finite size data. First one must numerically determine the intersection points of $\langle m_s^z(\tau)\rangle$ with the threshold value $m_{\mathrm{threshold}}$ required to calculate the finite size exponent $z(L,L')$ between two system sizes. Here we choose $L$ and $L'=2L$, defining $z(L)\equiv z(L,2L)$. To extract $\tau_r$ we fill a window around the threshold with QMC data and then use a polynomial (or some other appropriate function) to interpolate the data and numerically find the crossing point of the interpolation and $m_\mathrm{threshold}$. 

\begin{figure}[!h]
	\includegraphics[width=0.45\textwidth]{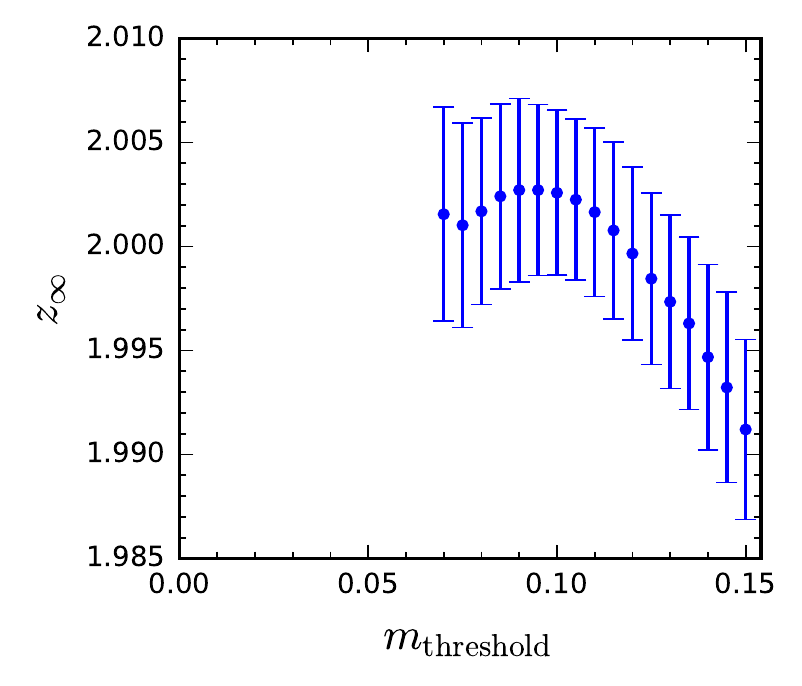}
	\caption{\label{fig:clean_exp} Extrapolated values of the asymptotic dynamic exponent $z_\infty$ of the clean 2D Heisenberg antiferromagnet as a function of the threshold value $m_\mathrm{threshold}$ used to perform the $\tau$-axis rescaling.}
\end{figure}

\begin{figure}
	\includegraphics[width=0.45\textwidth]{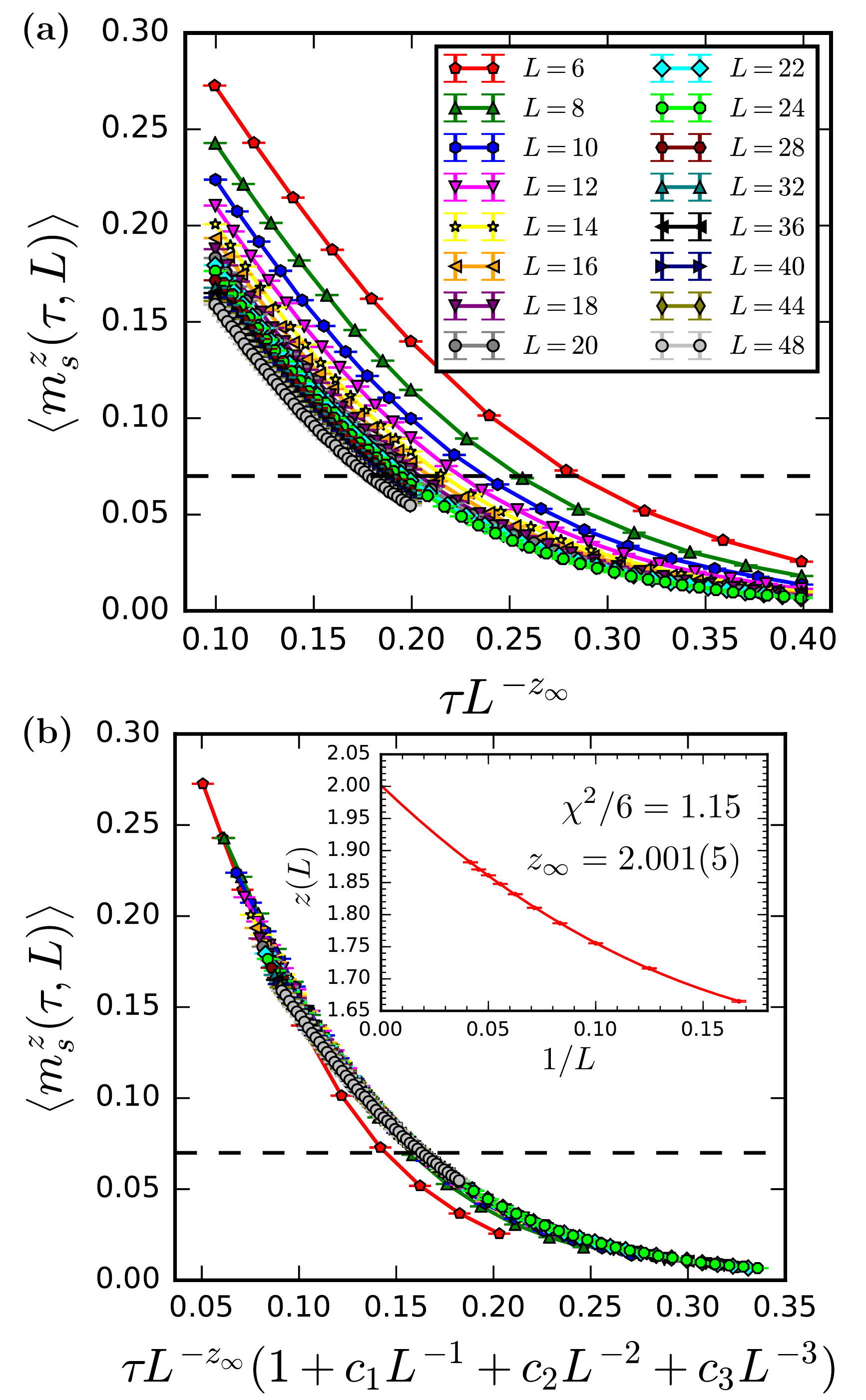}
	\caption{\label{fig:Clean_col} Scaling collapse of $\langle m_s^z(\tau)\rangle$ in the Heisenberg model on a square lattice without (a) and with (b) finite-size corrections to the leading scaling form $L^{-z_\infty}$. The dotted black lines denote the threshold value along which the curves are collapsed; $m_\mathrm{threshold}=0.07$. In (a), in the leading power-law $L^{-z_\infty}$ the exponent is $z_\infty=2.001$. In (b) the finite-size correction to $\Delta(L)$ are calculated from the fit to the size dependence of the exponent shown in the inset.}
\end{figure}

Next we must derive the finite size corrections to $z(L)$ which is dependent on the parameterization the finite size corrections in $\Delta(L)$. For this model it is natural to assume that $\Delta(L)$ should be an analytic function in $1/L$ because the model has no critical fluctuations. keeping terms up to order $L^{-3}$ in $\Delta(L)$ we find the finite size corrections to $z(L)$ are
\begin{multline}
z(L)=z_\infty+\frac{c_1}{2 \log (2)}\frac{1}{L}-\frac{3 \left(c_1^2-2 c_2\right)}{8 \log
   (2)}\frac{1}{L^2}\\+\frac{7 \left(c_1^3-3 c_2 c_1+3 c_3\right)}{24 \log (2)}\frac{1}{L^3}.\label{eq:Clean_fit}
\end{multline}
The extrapolated dynamic exponents are shown as a function $m_{\mathrm{threshold}}$ in Fig.~\ref{fig:clean_exp} (note that the values in Fig.~\ref{fig:clean_exp} are correlated because $m_\mathrm{threshold}$ values can be arbitrarily close to one another). The extrapolated values of  $z_\infty$ for low $m_\mathrm{threshold}$ are in excellent agreement with the analytic result of $z_\infty=2$. For higher values for $m_\mathrm{threshold}$ the disagreement is natural because at short times the many-body wavefunction still has an overlap with high-energy states. As the system evolves in imaginary-time (as $m_\mathrm{threshold}$ decreases), the overlap with these higher energy states decay and so their effects on the effective dynamic exponent vanish. The parameters $c_i$'s coming from the exponent flow give us the same coefficients which parameterize the finite size corrections of $\Delta(L)$. Figure \ref{fig:Clean_col}(a) focuses on the final relaxation time scale and more clearly shows the finite size corrections to $L^{-z}$. By including the finite size corrections calculated from the extrapolation, we find much better scaling collapse; see Fig.~\ref{fig:Clean_col}(b).

\section{Heisenberg Antiferromagnet on Fractal Clusters \label{sec:percolating}}

The second example has the same Heisenberg interactions with nearest neighbors as the previous section, but the boundaries are no longer periodic and spins on the sites are randomly removed from the lattice with probability $1-p$. This dilution is particularly interesting when it is set to the percolation point. It has been shown that at this dilution, the characteristics of the low energy states are different from that of the standard quantum rotor picture \cite{sandvik02,wang06,wang10,changlani13,ghosh15}. The volume of these clusters scale as $N_c\sim L^{D_f}$ where $D_f=91/48$ is the fractal dimension and $L$ is a linear size of the cluster \cite{stauffer94}. It has been conjectured that the low energy states of this model are dominated by a generalization of the ``dangling spin'' concept---local sub-lattice imbalance in a region of a given cluster \cite{wang06,wang10} where en effective moment forms due to the inability of spins to pair up in a bipartite manner. On a Bethe lattice geometry the same phenomena was studied using the DMRG method \cite{ghosh15}. In this study it was shown explicitly that there exists a set of low lying quasi-degenerate (QD) eigenstates which remain separated from the higher energy eigenstates by a finite size gap $\Delta_\mathrm{QD}$, which goes to 0 slower then the spacing between the QD states. It was conjectured that these QD states decouple from the bulk and because they are made up of power-law localized magnetic moments which interact with each other across the cluster, as had been previously deduced based on scaling behaviors of quantities probing the low-energy excitations indirectly \cite{wang06,wang10}. For this reason, diluted Heisenberg models have a larger dynamic exponent than predicted by the quantum rotor picture \cite{wang06,wang10,changlani13,ghosh15}.

Beyond the interesting physics of this model, the disorder should prove a more robust test of our scaling hypothesis and method for extracting the dynamic exponent. The major difference between this model and the last is the type of finite size corrections we see. In the unadulterated 2D case, the model is very far away from any sort of critical behavior and so the corrections are analytic in $L^{-1}$ but here we can not assume this as there are fluctuations driven by the classical percolation threshold \cite{sandvik02}. However, since we are only interested in extracting the dynamic exponent its perfectly reasonable to parameterize the finite size corrections to be analytic in $N^{-1}$, where $N$ is the number of sites making up a cluster. 

\begin{figure}
	\includegraphics[width=0.45\textwidth]{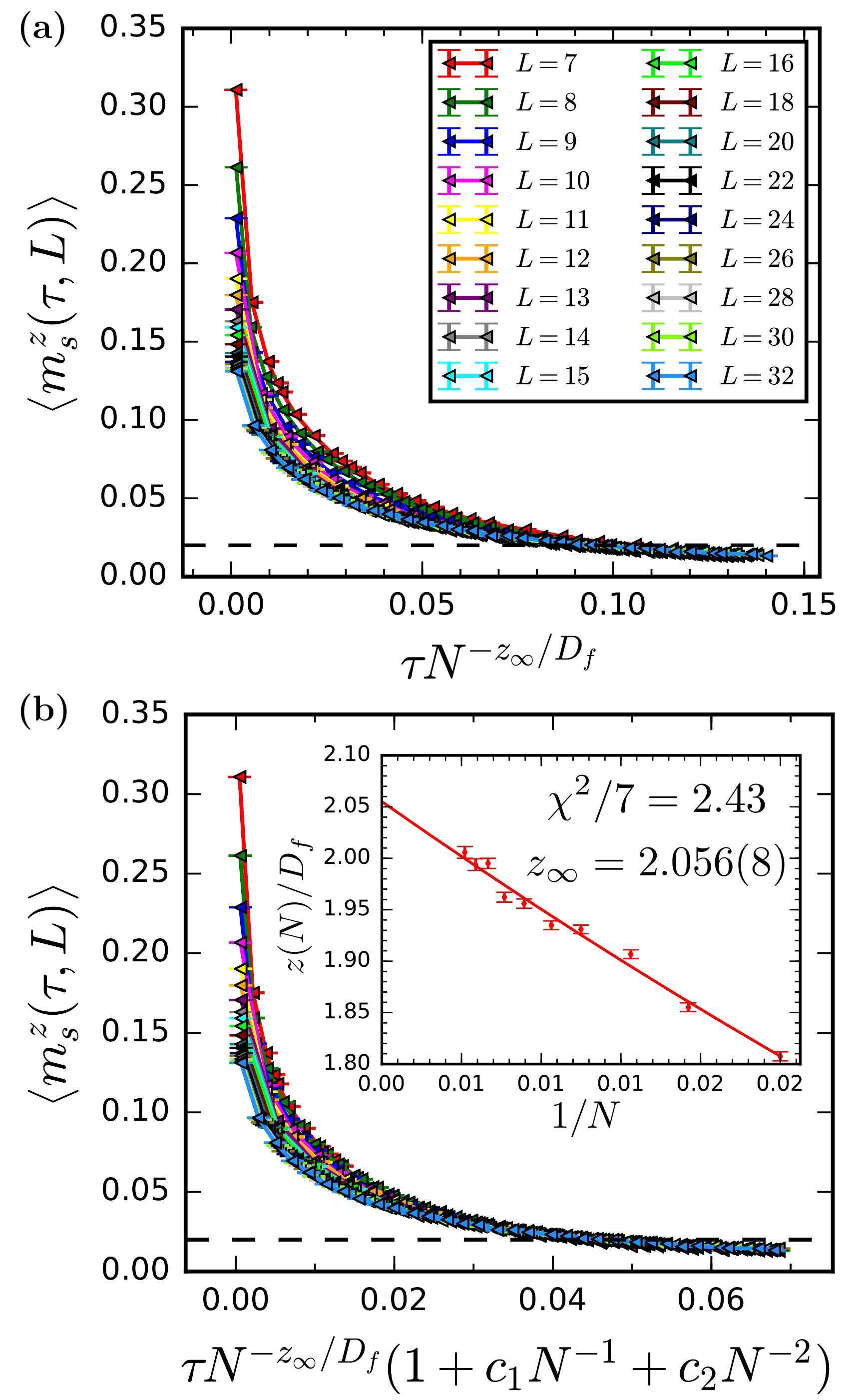}
	\caption{\label{fig:Cluster_col} 
	Finite-size scaling analysis of $\langle m_s^z(\tau)\rangle$ in the Heisenberg model on a percolating cluster (averaged over dilution realizations). The black dotted line denotes the threshold value along which the curves are collapsed; $m_\mathrm{threshold}=0.02$.  (a) shows the data collapse with the leading behavior $N^{-z_\infty/D_f}$ used for rescaling the $x$-axis, with $z_\infty/D_f=2.065$. (b) shows the scaling collapse including finite-size correction to $\Delta(N)$ corresponding to the fit of the finite-size flow shown in the inset.}
\end{figure}

To perform the dilution averaging we employ a procedure similar to what is outlined in Ref.~\cite{sandvik02}. Each cluster is constructed with a fixed number of sites $N_c=L^{D_f}$ where we round up to an even integer to insure that the ground state can have total $S=0$. The finite-size scaling of the low energy gap is implemented in the following form:
\begin{equation}
\Delta(N)= N^{-z_\infty/D_f}(1+c_1 N^{-1} + c_2 N^{-2}+\cdots). 
\end{equation}
The finite-size flow of the effective dynamic exponents are calculated between system sizes of length $L$ and $2L$, which implies $N'=2^{D_f}N$, for the cluster sizes. Keeping terms up to order $N^{-2}$ in finite size corrections to $\Delta(N)$ we obtain the following expansion for the finite-size effective dynamic exponent $z(N)$:
\begin{multline}
\frac{z(N)}{D_f}=\frac{z_\infty}{D_f}+\frac{c_1\left(2^{D_f}-1\right)}{2^{D_f} \log (2^{D_f})}\frac{1}{N}\\-\frac{3\left(2^{2D_f}-1\right) \left(c_1^2-2 c_2\right)}{2^{2D_f} \log
   (2^{D_f})}\frac{1}{N^2}.\label{eq:Cluster_fit}
\end{multline}
Results of a data-collapse analysis both with and without the scaling corrections are shown in Fig.~\ref{fig:Cluster_col}. 

\begin{figure}
	\includegraphics[width=0.45\textwidth]{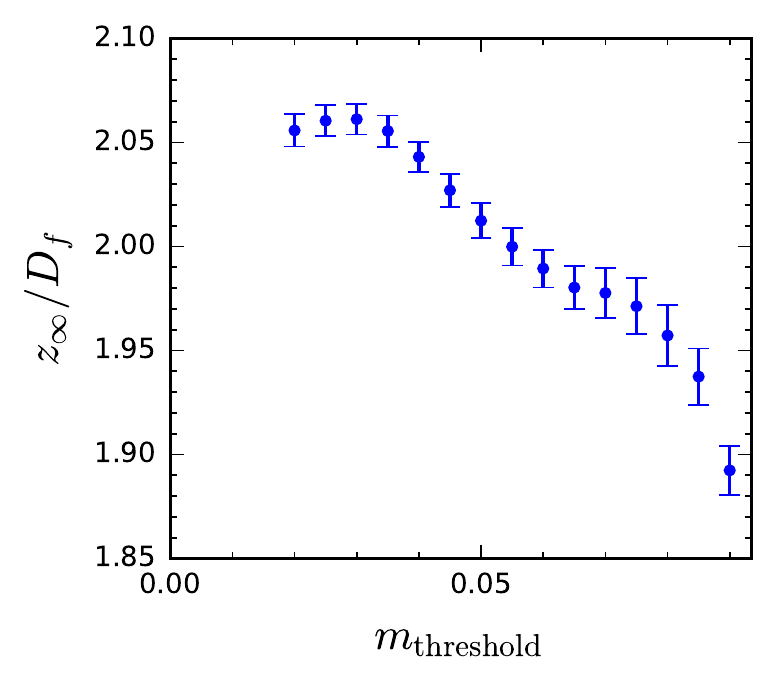}
	\caption{\label{fig:cluster_exp} 
	Extrapolated values of $z_\infty$ averaged over dilution realization for the site-diluted Heisenberg antiferromagnet as a function of the threshold value $m_\mathrm{threshold}$ used to perform the $\tau$-axis rescaling.}
\end{figure}

For our lowest value of $m_\mathrm{threshold}=0.02$, $z_\infty/D_f=2.056(8)$ or $z_\infty=3.90(1)$, slightly larger than what was found previous studies\cite{wang06,wang10}. Although we may have introduced some systematic error by assuming the finite size corrections decay as $1/N$, the inset in Fig.~\ref{fig:Cluster_col}(b) shows that for the largest system size ($N=714$), $z(N)/D_f \geq 2$. Because our system sizes are comparable to previous studies we know that this is not an issue of finite size effects. We also know that in imaginary-time evolution, the weights coming from an eigenstate $|n\rangle$ relaxes on a timescale of $\tau_n=1/(\epsilon_n-\epsilon_0)$, meaning that higher energy states always decay faster than low-energy states and therefore the effective dynamic exponents must be monotonically increasing as $m_\mathrm{threshold}\rightarrow 0$. This is consistent with results in the previous section and with the threshold dependence of the extrapolated dynamic exponent shown in Fig.~\ref{fig:cluster_exp}.

\section{Conclusion \label{summary}}

In summary we have shown that the relaxation of the order parameter in imaginary-time can be used to quantitatively extract low-energy properties of a given model. In particular, we have developed a scaling method which allows one to extract the dynamic exponent by performing scaling collapse of the order parameter along the imaginary-time axis. The imaginary-time evolution of the order parameter in a 2D Heisenberg antiferromagnets was used as a test-case to show that one can extract the correct dynamic exponent, which in this case is related to the Anderson tower of quantum rotor states. We also studied the relaxation of the order parameter of the corresponding site dilute Heisenberg antiferromagnet at its percolation threshold and found that the scaling theory gives a dynamic exponent even larger than seen in previous studies; $z=3.90$ versus $z\approx 3.7$ from Refs.~\cite{wang06,wang10}. These results conform with the notion that there is a set of low-energy states due to quasi-localized moments---``dangling spins'' and their generalizations to larger regions of local sublattice imbalance~\cite{wang06,wang10}.

As there are very clear numerical results that the ground state of the Heisenberg model on the percolating cluster does indeed have long range order~\cite{sandvik02}, there is still an open question if there is a quantum-rotor tower of states in these clusters, which presumably should be required for spontaneous breaking of the spin-rotation symmetry in the thermodynamic limit~\cite{anderson52,goldstone62}. If they do exist they would have relaxation times that would scale as $L^{D_f}$ according to the quantum rotor picture, however our method would not be able to isolate these states directly as it can only extract an effective dynamic exponent originating from the mixing of low and high energy states at short times.  The fact that we see a dynamic exponent with significant flow with the threshold value
of the mean order parameter in the system---much larger than in the unadulterated system---may in principle be an indication of the rotor states. However, the density of the rotor states should be much lower than the low-energy states arising from the quasi-localized moments. The drift in the exponent may therefore instead be related to a slowly changing dynamic exponent of these quasi-localized states as one goes to higher energies in their tower. We believe that this is the more likely scenario, with rotor states existing in the bulk of the cluster but not contributing significantly to the low-energy dynamics because of their much higher dynamic exponent and much lower density of states.

In this work we have only studied the spatially averaged order parameter but in the case of the percolating clusters it could be useful to look at the local order as a function of imaginary-time in a similar spirit as Refs.~\cite{wang06,wang10}. One would expect that if the localized spins are indeed decoupled from the bulk, that this would manifest itself as very different relaxation times between the localized moments and the bulk. One would even hope to see that the bulk relaxes to $0$ on time scales $\tau\sim L^{D_f}$ while the decoupled moments would relax on time scales of $L^{3.90}$. 

The QMC methods used here to simulate imaginary-time can be generalized to higher spin representations which, along with other methods, would give a more complete picture as to the low energy excitations of the higher spin versions of the Heisenberg antiferromagnets on percolating clusters. 

\section{Acknowledgments}

We would like to thank H. Changlani, A. Iaizzi, P. Patil, and A. Polkovnikov for useful discussions. The computational was performed on the Shared Computing Cluster administered by Boston University's Research Computing Services. This research was supported by the NSF under Grant No.~DMR-1410126.


\begin{thebibliography}{99}
\bibitem{paillard01}
M. Paillard, X. Marie, P. Renucci, T. Amand, A. Jbeli, and J. M. G\'erard, Phys. Rev. Lett. {\bf 86}, 1634 (2001).

\bibitem{kaiser14}
S. Kaiser, C. R. Hunt, D. Nicoletti, W. Hu, I. Gierz, H. Y. Liu, M. Le Tacon, T. Loew, D. Haug, B. Keimer, and A. Cavalleri, Phys. Rev. B {\bf 89}, 184516 (2014).

\bibitem{loth10}
S. Loth, M. Etzkorn, C. P. Lutz, D. M. Eigler, A. J. Heinrich, Science {\bf 329} 1628 (2010).

\bibitem{bloch05}
I. Bloch, Nat. Phys. {\bf 1}, 23 (2005).

\bibitem{sadler06}
L. E. Sadler, J. M. Higbie, S. R. Leslie, M. Vengalattore and D. M. Stamper-Kurn, Nature {\bf 443}, 312 (2006).

\bibitem{ritter07}
S. Ritter, A. \"Ottl, T. Donner, T. Bourdel, M. K\"ohl, and T. Esslinger Phys. Rev. Lett. 98, 090402 (2007).

\bibitem{trotzky11}
S. Trotzky, Y-A. Chen, A. Flesch, I. P. McCulloch, U. Schollwöck, J. Eisert and I. Bloch, Nat. Phys. {\bf 8}, 325 (2012).

\bibitem{schreiber15}
M. Schreiber, S. S. Hodgman, P. Bordia, H. P. L\"uschen, M. H. Fischer, R. Vosk, E. Altman, U. Schneider, I. Bloch, Science {\bf 349}, 842 (2015).

\bibitem{nandkishore15}
R. Nandkishore and D. A. Huse, Annu. Rev. Conden. Ma. P. A {\bf 6} 15 (2015).

\bibitem{vosk13}
R. Vosk and E. Altman, Phys. Rev. Lett. {\bf 110}, 067204 (2013).

\bibitem{pal10}
A. Pal and David A. Huse, Phys. Rev. B {\bf 82}, 174411 (2010).

\bibitem{oganesyan07}
V. Oganesyan and David A. Huse, Phys. Rev. B {\bf 75}, 155111 (2007).

\bibitem{imbrie16}
J. Z. Imbrie, J. Stat. Phys. {\bf 163}, 998 (2016).

\bibitem{gornyi05}
I. V. Gornyi, A. D. Mirlin, and D. G. Polyakov, Phys. Rev. Lett. {\bf 95}, 206603 (2005).

\bibitem{baskoa06}
D. M. Baskoa, I.L. Aleiner, B.L. Altshuler, Ann. Phys. {\bf 321}, 1126 (2006).

\bibitem{daley04}
A. J. Daley, C. Kollath, U. Schollwöck and G. Vidal5, J. Stat. Mech. P04005 (2004).

\bibitem{schollwock05}
U. Schollwöck, Rev. Mod. Phys. {\bf 77} 259 (2005).

\bibitem{schollwock11}
U. Schollwöck, Ann. Phys. {\bf 326}, 96 (2011).

\bibitem{jarrell96}
M. Jarrell and J. E. Gubernatis, Phys. Rep. {\bf 269}, 133 (1996).

\bibitem{degrandi11}
C. De Grandi, A. Polkovnikov, and A. W. Sandvik, Phys. Rev. B {\bf 84}, 224303 (2011).

\bibitem{degrandi13}
C. D. Grandi, A. Polkovnikov, and A. W. Sandvik, J. Phys.: Cond. Mat. {\bf 25}, 404216 (2013).

\bibitem{liu13}
C.-W. Liu, A. Polkovnikov, and A. W. Sandvik, Phys. Rev. B {\bf 87}, 174302 (2013).

\bibitem{kibble76}
T. W. B. Kibble, J. Phys. A {\bf 9}, 1387 (1976).

\bibitem{zurek85}
W. H. Zurek, Nature (London) {\bf 317}, 505 (1985).

\bibitem{polkovnikov05}
A. Polkovnikov, Phys. Rev. B {\bf 72}, 161201(R) (2005).

\bibitem{zurek05}
W. H. Zurek, U. Dorner, and P. Zoller, Phys. Rev. Lett. {\bf 95}, 105701 (2005).

\bibitem{dziarmaga05}
J. Dziarmaga, Phys. Rev. Lett. {\bf 95}, 245701 (2005).

\bibitem{zhong05}
F. Zhong and Z. Xu, Phys. Rev. B {\bf 71}, 132402 (2005).

\bibitem{chandran12}
A. Chandran, A. Erez, S. S. Gubser, and S. L. Sondhi, Phys. Rev. B {\bf 86}, 064304 (2012).

\bibitem{liu14}
C.-W. Liu, A. Polkovnikov, and A. W. Sandvik, Phys. Rev. B {\bf 89}, 054307 (2014).

\bibitem{micheal14}
M. Kolodrubetz, Phys. Rev. B {\bf 89}, 045107 (2014).

\bibitem{yin14}
S. Yin, P. Mai, and F. Zhong, Phys. Rev. B {\bf 89}, 144115 (2014).

\bibitem{zhang14}
S. Zhang, S. Yin, and F. Zhong, Phys. Rev. E {\bf 90}, 042104 (2014).

\bibitem{shu17}
Y.-R. Shu, S. Yin, D.-X. Yao, arXiv:1705.05931.

\bibitem{chiocchetta16}
A. Chiocchetta, A. Gambassi, S. Diehl, and J. Marino, Phys. Rev. B {\bf 94}, 174301 (2016).

\bibitem{liu16}  
W. Liu and U. C. Tuber, J. Phys. A: Math. Theor. {\bf 49}, 434001 (2016).

\bibitem{shao15}
H. Shao, W. Guo, and A. W. Sandvik, Phys. Rev. B {\bf 91}, 094426 (2015).

\bibitem{anderson52}
P. W. Anderson, Phys. Rev. {\bf 86 }, 694 (1952).

\bibitem{wang06}
L. Wang and A. W. Sandvik, Phys. Rev. Lett. {\bf 97}, 117204 (2006).

\bibitem{wang10}
L. Wang and A. W. Sandvik, Phys. Rev. B {\bf 81}, 054417 (2010).

\bibitem{ghosh15}
S. Ghosh, H. J. Changlani, and C. L. Henley, Phys. Rev. B {\bf 92}, 064401 (2015).

\bibitem{goldstone62}
J. Goldstone, A. Salam, and S. Weinberg, Phys. Rev. {\bf 127}, 965 (1962).

\bibitem{vojta95}
T. Vojta and J. Schmalian, Phys. Rev. Lett. {\bf 95}, 237206 (2005).

\bibitem{sandvik02}
A. W. Sandvik, Phys. Rev. B {\bf 66}, 024418 (2002).

\bibitem{d_fisher89}
D. S. Fisher, Phys. Rev. B {\bf 39}, 11783 (1989).

\bibitem{nightingale82}
P. Nightingale Journal of Applied Physics {\bf 53}, 7927 (1982).

\bibitem{brezin82}
E. Brezin, Journal de Physique {\bf 43}, 15 (1982).

\bibitem{derrida83}
L. D. S. B. Derrida, Journal de Physique {\bf 43}, 475 (1983).

\bibitem{luck85}
M. Luck, Phys. Rev. B {\bf 31}, 3069 (1985).

\bibitem{neuberger89}
H. Neuberger and T. Ziman, Phys. Rev. B {\bf 39}, 2608 (1989).

\bibitem{hasenfratz93}
P. Hasenfratz and F. Niedermayer, Z. Phys. B {\bf 92}, 91 (1993).

\bibitem{sandvik10}
A. W. Sandvik, AIP Conf. Proc. {\bf 1297},  135 (2010).

\bibitem{sandvik10_2}
A. W. Sandvik and H. G. Evertz Phys. Rev. B {\bf 82}, 024407 (2010).

\bibitem{lavalle98}
C. Lavalle, S. Sorella, and A. Parola, Phys. Rev. Lett. {\bf 80}, 1746 (1998).

\bibitem{sandvik97}
A. W. Sandvik, Phys. Rev. B {\bf 56} 11678 (1997).

\bibitem{changlani13}
H. J. Changlani, S. Ghosh, S. Pujari, and C. L. Henley, Phys. Rev. Lett. {\bf 111}, 157201 (2013).

\bibitem{stauffer94}
D. Stauffer and A. Aharony, {\it Introduction to percolation theory} (CRC press, 1994).	



\end{thebibliography}
\end{document}